\documentclass[twocolumn]{aastex6}
\usepackage{natbib}
\usepackage{amsmath}
\usepackage{breakurl}
\usepackage{hyperref}
\usepackage{textcomp}
\usepackage{soul}

\newcommand\blfootnote[1]{%
  \begingroup
  \renewcommand\thefootnote{}\footnote{#1}%
  \addtocounter{footnote}{-1}%
  \endgroup
}

\begin{document}

\title{CAPABILITIES OF EARTH-BASED RADAR FACILITIES FOR NEAR-EARTH ASTEROID OBSERVATIONS}

\author{Shantanu. P. Naidu\altaffilmark{1}, Lance. A. M. Benner\altaffilmark{1}, Jean-Luc Margot\altaffilmark{2}, Michael. W. Busch\altaffilmark{3}, Patrick. A. Taylor\altaffilmark{4}}

\altaffiltext{1}{Jet Propulsion Laboratory, California Institute of Technology, Pasadena, CA 91109-8099, USA}
\altaffiltext{2}{Department of Earth, Planetary, and Space Sciences, University of California, Los Angeles, CA 90095, USA}
\altaffiltext{3}{SETI Institute, Mountain View, CA, 94043, USA}
\altaffiltext{4}{Arecibo Observatory, Universities Space Research Association, Arecibo, Puerto Rico, 00612, USA}

\begin{abstract}
  
  We evaluated the planetary radar capabilities at Arecibo, the
  Goldstone 70-m DSS-14 and 34-m DSS-13 antennas, the 70-m DSS-43
  antenna at Canberra, the Green Bank Telescope, and the Parkes Radio
  Telescope in terms of their relative sensitivities and the number of
  known near-Earth asteroids (NEAs) detectable per year in monostatic
  and bistatic configurations. In the 2015 calendar year, monostatic
  observations with Arecibo and DSS-14 were capable of detecting 253
  and 131 NEAs respectively, with signal-to-noise ratios (SNRs)
  greater than 30/track. Combined, the two observatories were capable
  of detecting 276 NEAs. Of these, Arecibo detected 77 and Goldstone
  detected 32, or 30\% and 24\% the numbers that were possible. The
  two observatories detected an additional 18 and 7 NEAs respectively,
  with SNRs of less than 30/track. This indicates that a substantial
  number of potential targets are not being observed. The bistatic
  configuration with DSS-14 transmitting and the Green Bank Telescope
  receiving was capable of detecting about 195 NEAs, or $\sim50$\%
  more than with monostatic observations at DSS-14. Most of the
  detectable asteroids were targets of opportunity that were
  discovered less than 15 days before the end of their observing
  windows. About 50\% of the detectable asteroids have absolute
  magnitudes $> 25$, which corresponds diameters $< \sim$30 m.
  
\end{abstract}

\blfootnote{\textcopyright~2016. All rights reserved.}
\section{Introduction}

Ground-based radar is an invaluable tool for improving the orbits and
characterizing the physical properties of near-Earth asteroids
(NEAs). Such characterization is essential for identifying and
mitigating impact threats, planning robotic and human space missions,
and for advancing our scientific understanding of asteroids. Radar
range and Doppler measurements of NEAs often have fractional precision
better than 1 in $10^7$, which can provide dramatic improvements to
asteroid orbits, prevent loss of newly discovered objects, and
increase the window of reliable predictions of trajectories by decades
to centuries~\citep{ostro04}. In some cases, signal-to-noise ratios
(SNRs) are high enough to obtain delay-Doppler images of the objects
with range resolutions as fine as 1.875 m. Delay-Doppler images
provide a powerful technique to see surface features, obtain shape
models~\citep[e.g.,][]{naidu13}, discover natural
satellites~\citep[e.g.,][]{margot02}, and estimate masses and
densities from binary systems~\citep[e.g.,][]{ostro06,
  naidu15b,margot16} and from non-gravitational accelerations due to
the Yarkovsky effect~\citep[e.g.,][]{benner16,vokrouhlicky16}. Radar
observations also help support spacecraft missions such as
NEAR-Shoemaker, Hayabusa, Chang'e 2, EPOXI, OSIRIS-REx, and the
Asteroid Redirect Mission (ARM).

In 2005, the United States Congress passed the George E. Brown,
Jr. Act that directed NASA to detect, track, and characterize
near-Earth objects larger than 140 m in diameter. The objectives were
based on a National Aeronautics and Space Administration (NASA) report
(Near-Earth Object Science Definition Team 2003) which concluded that
such objects are capable of penetrating through Earth's atmosphere and
causing regional destruction in an impact. In 2010, the goals of the
George E. Brown, Jr. act were incorporated in the National Space
Policy of the United States of America
(\url{https://www.whitehouse.gov/sites/default/
files/national_space_policy_6-28-10.pdf})
that guides the NASA administrator to “pursue capabilities, in
cooperation with other departments, agencies, and commercial partners,
to detect, track, catalog, and characterize near-Earth objects to
reduce the risk of harm to humans from an unexpected impact on our
planet and to identify potentially resource-rich planetary objects.”

\begin{figure}
    \plotone{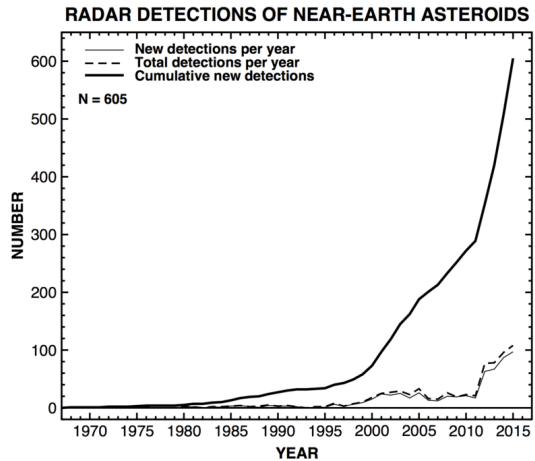}

  \caption{Total number of near-Earth asteroids detected by
    ground-based radar facilities each year. The cumulative number of
    radar detections as a function of time is shown using a solid bold
    line.}
  
  \label{fig:radardetect}
\end{figure}

The 305-m Arecibo Observatory in Puerto Rico (2380 MHz, 12.6 cm) and
the 70-m DSS-14 antenna at the Goldstone Deep Space Communications
Complex in California (8560 MHz, 3.5 cm) are the only telescopes with
radar transmitters that are regularly used to observe NEAs.
\textbf{Arecibo and DSS-14 have observed hundreds of NEAs since the
  1970s (Figure~\ref{fig:radardetect}).}

In 2014, the 34-m DSS-13 antenna at Goldstone was equipped with an 80
kW transmitter that operates at a frequency of 7190 MHz (4.2 cm
wavelength). This transmitter
allows delay-Doppler
imaging with range resolutions up to 1.875 m, which is twice as fine
as the finest range resolution available at DSS-14
and four times finer than the finest resolution
available at Arecibo.
The transmitter at DSS-13 was used to image two NEAs in 2015.

The 100-m Green Bank Telescope (GBT) is occasionally used in a
bistatic configuration to receive radar echoes from NEAs, with DSS-14,
DSS-13, or Arecibo transmitting. The GBT does not have a radar
transmitter.

In November 2015, the 70-m DSS-43 antenna in Canberra \textbf{(2290 MHz,
13.1~cm)} and the 64-m Parkes Radio Telescope obtained the first radar
detection of a NEA in Australia with (413577) 2005 UL5.  This is an
important capability because these telescopes are located in the
southern hemisphere and can point to high southern declinations that
cannot be seen by Arecibo, Goldstone, or the GBT. The bistatic DSS-43
to Parkes configuration is currently the most sensitive option for
radar observations of NEAs at high southern declinations. Because
DSS-43 and Parkes are offset substantially in longitude from Arecibo
and Goldstone, they can also allow observations of asteroids at times
when targets are below the horizon at the other radar facilities,
which could extend rotational coverage for some objects and allow
radar observations in Australia that are not possible elsewhere due to
scheduling conflicts.

\textbf{Delay-Doppler images are important data products of radar
  observations and resolve the target in time delay and Doppler
  frequency (See \citet{ostro93} for details).  The resolution along
  the delay axis depends on the transmitted signal and cannot be
  changed once the data is recorded.  The finest possible delay
  resolution is limited by the maximum transmitter bandwidth.}

\textbf{The choice of Doppler resolution is more flexible as it
  depends on the Fourier Transform length, which is selected while
  processing the data and is limited by the integration time. The
  finest possible Doppler resolution is given by the inverse of the
  total integration time. For monostatic observations this is
  approximately equal to the round-trip light time to the target
  (minus the time required to switch from transmitter to
  receiver). With bistatic observations, it is limited by the length
  of continuously recorded data.  During observations, the delay and
  Doppler resolutions are subjectively chosen such that the
  corresponding spatial scales along each axis is roughly equal. Other
  factors are also considered but are outside the scope of this
  discussion.  }

In this paper, we explore the capabilities of these ground-based
radars in terms of their relative sensitivities and the number of NEAs
detectable per year in various monostatic and bistatic
configurations. We seek to answer the question: If resources were not
an issue, how many NEAs could be observed with current radar
facilities per year? This study is important for strategic planning of
observations, implementing policies, and future upgrades.

The first study into this topic was by \citet{jurgens77}, who found
that 60 out of the 1984 then-known asteroids (near-Earth and main
belt) were detectable at either Goldstone or Arecibo over a period of
ten years between 1977 and 1987. The number of known asteroids has
grown tremendously since then and the number of objects detectable by
these facilities is now dramatically higher. About 605 had been
detected through the end of 2015 (Figure~\ref{fig:radardetect}).

\citet{giorgini08} performed a study to determine the
improvements in number of detectable asteroids, trajectory
predictions, and physical characterization of asteroids achieved by
doubling the transmitter power at DSS-14, adding a southern hemisphere
planetary radar, and increasing the transmitter bandwidth of the
transmitter. The study used a simulated NEA population to estimate the
improvements.

In 2010, the National Research Council, at the request of NASA,
conducted a study that found that planetary radar plays a crucial role
for achieving the goals of the George E. Brown, Jr. Act (Shapiro et
al., 2010). They estimated that about 410 near-Earth objects could be
observed by radar (Arecibo \& Goldstone) in a 1-year interval starting
in 2008 May. Out of these, 140 had been discovered before 2008 May and
270 were found during the time interval considered in the report.

During calendar year 2015, more than 1500 NEAs were discovered
(\url{http://minorplanetcenter.net/iau/lists/YearlyBreakdown.html}), the
largest number of NEA discoveries in a single year to date. The NEA discovery
rate is expected to rise in the next few years as new survey
telescopes such as Pan-STARRS 2 and ATLAS become fully operational and
new wider-field cameras are installed at the Catalina Sky Survey (CSS)
telescopes (Christensen et al., 2015). Because many NEAs are
detectable by radar during their discovery apparition~\citep{ostro04},
2015 provides the most accurate indication of the number of
radar-detectable NEAs under current circumstances. Often the discovery
apparition provides the best opportunity for radar observations for
decades~\citep{ostro04} and considerable effort is devoted to
observe such targets of opportunity (TOOs). For example, in 2015, 45
TOOs were observed with Arecibo and 15 TOOs were observed with DSS-14.

\section{Methods}
\label{sec:methods}
We performed an automated search of all NEAs listed on the Minor
Planet Center (MPC) website
(\url{http://www.minorplanetcenter.net/iau/lists/MPLists.html}) that
were discovered before the end of 2015 to identify all radar observing
opportunities in that year. We downloaded trajectories of the NEAs
from the Jet Propulsion Laboratory's Horizons ephemeris service
(\url{http://ssd.jpl.nasa.gov/?horizons_doc}). We estimated radar SNRs
for transmission/reception at Arecibo, DSS-13, DSS-14, DSS-43, Parkes,
and the GBT for this study. For each telescope, we created a short
list of objects that came within 1 astronomical unit (au) of Earth and
passed through the declination window of the telescope
(Table~\ref{tab:parameters}). In this step, we computed the positions
of the objects at 1-hour intervals throughout 2015. For each object in
the short-list, detailed observing windows, or radar track durations,
were computed by finding the time intervals in which the asteroid was
above the minimum viewable elevation of the telescope
(Table~\ref{tab:parameters}).

For each observing window we computed the minimum range to the target
and estimated the SNR of its radar echo as $P_{rx}/\Delta P_{noise}$,
where $P_{rx}$ is the power of the received echo and $\Delta
P_{noise}$ is the standard deviation of the receiver noise. $P_{rx}$
can be estimated using the radar equation~\citep[e.g.,][]{ostro93},

\begin{equation}
  P_{rx}=\frac{P_{tx}G_{tx}G_{rx}\lambda^2\sigma}{(4\pi)^3 R^4}.
  \label{eq:radareq}
\end{equation}

Here, $P_{tx}$ is the transmitter power, $G_{tx}$ and $G_{rx}$ are the
gain of the transmitting and receiving antennae, $\lambda$ is the
radar wavelength, $\sigma$ is the radar cross-section of the target,
and $R$ is the distance between the target and the reference point of
the antenna. For monostatic radar observations, the receiving and
transmitting antennae are the same so $G_{tx} = G_{rx}$. Antenna gain,
$G$, is given by $4\pi\eta A_{ant}/\lambda^2$, where $\eta$ is the
aperture efficiency of the antenna and $A_{ant}$ is the geometric area
of the antenna.

$\Delta P_{noise}$ is given by

\begin{equation}
  \Delta P_{noise} = \frac{k T_{sys} \Delta f}{(\Delta t \Delta f)^\frac{1}{2}}.
  \label{eq:noise}
\end{equation}

Here $k$ is Boltzmann's constant, $T_{sys}$ is the receiver
temperature, $\Delta f$ is the frequency resolution, and $\Delta t$ is
the total integration time of the received signal. The SNR is
maximized if $\Delta f$ is equal to the bandwidth ($B$) of the
echo. Because receiver noise is stochastic in nature, the standard
deviation of noise power falls off as the square root of the
integration time.

The radar cross-section and echo bandwidth depend on the physical
properties of the target and are often unknown. The radar cross
section is given by $\sigma=\hat{\sigma}A$, where $\hat{\sigma}$ is
the radar albedo and $A$ is the projected area of the target. For a
spherical object, the echo's bandwidth is given by $B=4\pi D
\cos{\delta}/\lambda P$, where $D$ is the diameter, $\delta$ is the
sub-radar latitude of the object, $\lambda$ is the radar wavelength,
and $P$ is the spin period of the object. The telescope parameters
used for the calculations are provided in Table 1. The relative
sensitivities per round-trip light time (RTT), the time for the
transmitted signal to reach, reflect off of, and return from the
target, are given in Table~\ref{tab:relsens} for various combinations
of transmitters and receivers.

\begin{deluxetable*}{lllllll}
  \tablecaption{Telescope parameter assumptions\label{tab:parameters}}
  \tablehead{\colhead{} & \colhead{DSS-13} & \colhead{DSS-14} & \colhead{DSS-43} & \colhead{Arecibo} & \colhead{GBT} & \colhead{Parkes}}
  \startdata
  Declination range & $-35^\circ$ to $+90^\circ$ & $-35^\circ$ to $+90^\circ$ & $-90^\circ$ to $+34.5^\circ$ & $-1^\circ$ to $+38^\circ$ & $-46^\circ$ to $+90^\circ$ & $-90^\circ$ to $+26.5^\circ$ \\
  Min. elevation (degrees) & 20                 & 20                       & 20                          & 70                      & 5                         & 30.5                        \\
  Accessible sky fraction (\%) & 79             & 79                       & 78                          & 32                      & 86                        & 72                          \\
  Diameter (m)                 & 34             & 70                       & 70                          & 305                     & 100                       & 64                          \\
  Aperture efficiency          & 0.71           & 0.64                     & 0.64                        & 0.38                    & 0.71                      & 0.45                        \\
                               &                &                          &                             & 0.17 (X-band)           &                           &                             \\
  Transmitter frequency (MHz)  & 7190           & 8560                     & 2290                        & 2380                    & NA                        & NA                          \\
  Transmitter power (kW)       & 80             & 450                      & 100                         & 900                     & NA                        & NA                          \\ 
  System temperature (K)       & 20             & 18                       & NA                          & 23                      & 25                        & 28                          \\
  \textbf{Transmit-receive switch time (s)} & NA         & 5                        & NA                          & 5                       & NA                        & NA                          \\
  \enddata
\end{deluxetable*}

\begin{deluxetable}{lll}
 \tablecaption{Relative sensitivities of various transmitter-receiver
   combinations\label{tab:relsens}}
 \tablehead{\colhead{Transmitter} & \colhead{Receiver} & \colhead{Relative sensitivity}}
 \startdata
 DSS-14 & DSS-14  & 1\\
        & Arecibo & 5.1\\
        & GBT     & 2.3 \\
        & DSS-13  & 0.3\\
 Arecibo & Arecibo & 15\\
         & GBT     & 5\\
         & DSS-13  & 0.6\\
         & DSS-14  & 2.2\\
 DSS-43   & Parkes  & 0.007\\
 DSS-43 (400 kW) & Parkes & 0.03 \\
 DSS-13   & Arecibo & 0.2\\
          & GBT     & 0.08\\
 \enddata
 
 \tablecomments{The first column indicates transmitting telescope, the
   second column indicates receiving telescope, and the third column
   indicates SNR/RTT values normalized to those at DSS-14. Transmitter
   and receiver parameters are provided in
   Table~\ref{tab:parameters}. }
\end{deluxetable}

We used the European Asteroid Research Node (\url{earn.dlr.de})
database, which is the most thorough and up-to-date database of NEA
physical properties, to obtain estimates of $D$ and $P$. When a
diameter estimate was not available, we computed a value for $D$ from
the absolute magnitude of the object by assuming a typical S-class
optical albedo of 0.18. For objects with unknown rotation periods, we
assumed a rotation period of 2.1 hours if the object was larger than
140 m in diameter and 0.5 hours for smaller objects. Spin periods for
the vast majority of asteroids greater than 140 m exceed 2.1
hours~\citep{pravec02}, so this threshold places a conservative lower
bound on the SNRs. Smaller asteroids exhibit a wide range of spin
periods, both faster and slower than 2.1 hours, and \textbf{0.5 hours
  is close to the median (See Figure 5 in \citet{harris16})}.

For bistatic observations, the radar signal can be transmitted for the
entire duration of the observing window (100\% duty cycle). For
monostatic observations, the same antenna transmits and receives so
the duty cycle of transmission cannot exceed 50\% and the total
integration time of the radar echo cannot be more than half the time
of the observing window. Our calculations also took into consideration
the finite time required to switch between the transmitter and
receiver, which makes the integration time several seconds less than
50\% of the observing window.

We adopted an SNR cutoff of $\geq 30$/track for the final list of
radar-detectable NEAs. Objects are detectable at lower estimated SNRs
but the rate of successful detection is close to 100\% above this
threshold.

Monthly ``survey nights'' are scheduled at Arecibo to observe as many
detectable NEAs (mostly TOOs) as possible in a nominally 8-h block of
observing time. During these nights the observers often target NEAs
with SNRs weaker than 30/track. This is tractable because Arecibo
radar tracks for single objects are usually short (less than 3 hours)
because the telescope can only point within 20 degrees of zenith. If a
target is not detected rapidly, a new target, if available, is
selected.

Asteroids with SNRs $< 30$/track are infrequently scheduled at
Goldstone because telescope is heavily subscribed with spacecraft
communications and only asteroids with high probabilities of success
are scheduled. Radar tracks at Goldstone are usually much longer than
at Arecibo and can, at least in principal, last more than 24 hours for
objects at high northern declinations. This implies that it may take
an unreasonably long time to achieve low SNRs at Goldstone for some
objects and it is not be practical to schedule such long tracks.

We assume that Canberra and Parkes would only be used for crucial
targets that are not visible at other radar facilities so we used a
lower threshold of SNR $\geq 15$/track for these telescopes.

\begin{figure}
  \plotone{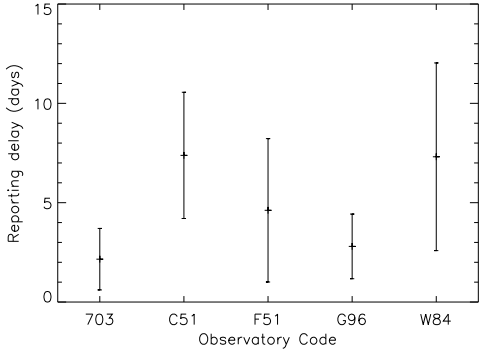}
  
  \caption{Means and standard deviations of discovery report delays
    for the top 5 observatories in terms of the number of NEAs
    discovered. Observatory codes designated by the MPC and sample
    sizes for each observatory are as follows: 703 – Catalina Sky
    Survey (142), C51 – NEOWISE mission (30), F51 – PanSTARRS I (320),
    G96 – Mt. Lemmon Survey (263), and W84 – Dark Energy Cam (35).}
  
\label{fig:reporting}  
\end{figure}

Our calculations account for the discovery announcement dates of the
radar-observable NEAs and excluded targets that were detectable by
radar only before their discoveries were announced. For targets
discovered by the Catalina Sky Survey (CSS) and Mt. Lemmon Survey we
added a typical reporting lag time of 2 days between discovery
observation and discovery announcements. For the Panoramic Survey
Telescope \& Rapid Response System (Pan-STARRS) we added 4 days, and
for all other observatories we adopted reporting lag times of 10
days. These typical reporting lag times were estimated by averaging
over 790 randomly selected asteroid discovery announcements from 2014
from the MPC website. Figure~\ref{fig:reporting} shows the means and
standard deviations of the reporting lag times for various
observatories. In this paper, we refer to asteroids that were
detectable by radar irrespective of their discovery dates as
``potentially detectable''. We use the term ``detectable'' to refer to
potentially detectable asteroids whose discoveries were announced
before the radar view periods ended.

\textbf{Radar beam sizes, approximately given by $\lambda/D_{ant}$,
  where $D_{ant}$ is the antenna diameter, are on the order of 1
  arcminute}, so pointing uncertainties should ideally be
significantly smaller than that prior to a radar observation. In
practice, we prefer $3\sigma$ pointing uncertainties $< 20$
arcseconds. Radar is not an efficient method for blindly searching for
asteroids due to its narrow beam width, the potentially enormous
Doppler shifts caused by the object's translational motion, and the
cost of the observations. Some known NEAs have large plane-of-sky
pointing, Doppler, and time delay uncertainties. For these objects we
ignored the uncertainties and adopted the nominal orbit. We often
request optical astrometry for such objects so only a few scheduled
objects are not detected anually due to large pointing
uncertainties. For some of the observable objects there is
insufficient time for scheduling and planning observations before the
objects exit the radar windows. More time is required to schedule a
target at Goldstone than at Arecibo because obtaining transmit
authorization from various government agencies is necessary due to
airspace restrictions. Other limitations to schedule NEAs for radar
observations include scheduling conflicts with other observing
projects, spacecraft communication, equipment problems, and
insufficient budget for staffing telescopes to observe all detectable
NEAs. DSS-43 and Parkes are currently not configured to observe TOOs
on short notice. These limitations make the number of observed targets
much lower than the number of detectable asteroids listed in this
paper. For Arecibo, DSS-14, and GBT we compared the numbers of
observable NEAs with the number of NEAs actually observed.

\section{Results}

\subsection{Arecibo monostatic}
\begin{figure}

  \plotone{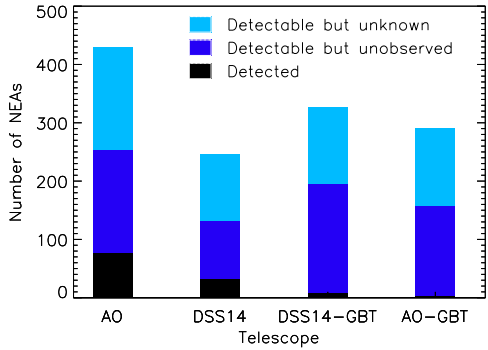}
  \caption{Histogram of currently known NEAs detectable and detected
    by various configurations of radar telescopes in 2015. AO stands
    for Arecibo Observatory and DSS-14 is the 70-m antenna at
    Goldstone. Each bar is divided into three parts: black indicates
    the asteroids that were observed by radar whereas light and dark
    blue colors represent asteroids that were not observed by
    radar. Asteroids in light blue were unknown during their radar
    observing windows. Dark blue represents asteroids that were known
    during their radar windows but not observed by radar.}
  \label{fig:mainhist}
\end{figure}

\begin{figure}
  \plotone{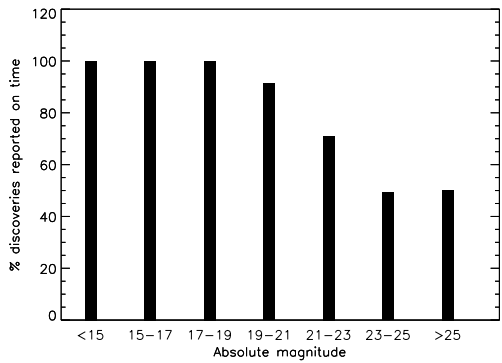}

  \caption{Fraction of potentially detectable asteroids at Arecibo for
    which discoveries were announced before their radar view periods
    with SNRs/track $> 30$ ended. }
  
  \label{fig:discvmag}
\end{figure}

We found 430 known and new objects that were potentially detectable at
Arecibo in 2015 with SNRs greater than 30/track
(Figure~\ref{fig:mainhist}). There were 253 discoveries ($\sim59\%$ of
the potentially detectable NEAs) that were announced by the Minor
Planet Center before radar view periods with SNR/track $\geq 30$
ended. The fraction of discoveries announced in time for radar
observations varies as a function of the absolute magnitudes of the
asteroids and is shown in
Figure~\ref{fig:discvmag}. \textbf{Discoveries of all potentially
  detectable asteroids brighter than absolute magnitude of 19
  (corresponding to diameters $>\sim500$ m) were announced before
  their radar view periods ended.}  Arecibo detected 77 NEAs in 2015
with estimated SNRs greater than 30/track, or about 30\% of the NEAs
considered as detectable in this study. An additional 18 NEAs were
detected with weaker SNRs.

\begin{figure}
  \plotone{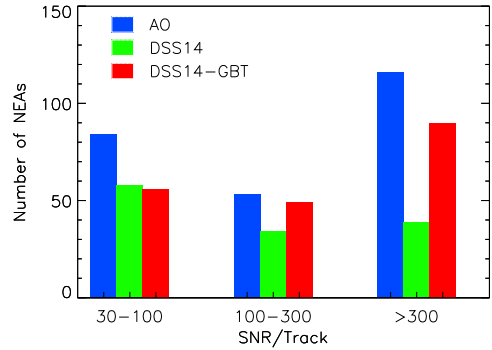}
  \caption{Number of radar-detectable asteroids in the low, medium,
    and high SNR categories for monostatic observations at Arecibo and
    DSS-14, and bistatic observations with DSS-14 and Green Bank.}
  \label{fig:snrbins}
\end{figure}

We divided the population of asteroids detectable by Arecibo into
three categories depending on the SNRs/track
(Figure~\ref{fig:snrbins}). The first category consists of asteroids
with SNRs/track between 30 and 100 for which we can typically obtain
ranging observations and echo power spectra that resolve them in
Doppler frequency. There are 84 detectable asteroids in this category
(18 were detected). The second category consists of targets with
SNRs/track between 100 and 300 that can typically yield delay-Doppler
images with coarse to medium resolutions (a few to a few tens of delay
pixels over the object). There are 53 detectable asteroids in this
category (23 were detected). The third category of targets have
SNRs/track $>300$. These are imaging targets that can be resolved with
range resolutions as fine as 7.5 m/pixel, the highest range resolution
available at Arecibo. There are 116 detectable asteroids in this
category (36 were detected).

\begin{figure*}
  \plotone{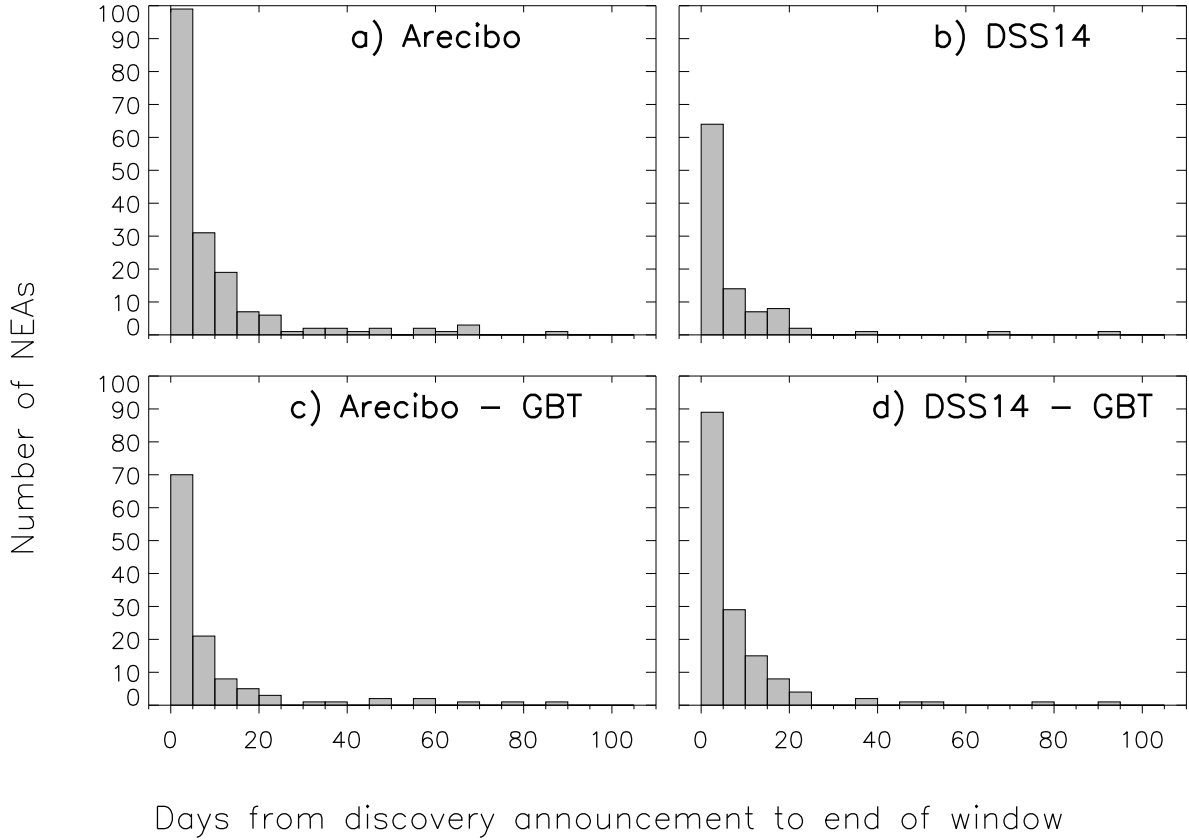}

  \caption{Histograms showing the number of detectable NEAs as a
    function of the time between discovery announcement to the end of
    observing window for Arecibo, DSS-14, Arecibo-GBT and DSS-14-GBT.}
  
  \label{fig:notice}
\end{figure*}

For TOOs that were discovered within 100 days of their radar view
periods, Figure~\ref{fig:notice} shows a histogram of the number of
days between discovery announcements and the ends of the observing
windows for the corresponding objects.
Almost all
of new discoveries
are detectable by radar only within 15 days of the discovery
announcement. If the delay in reporting NEA discoveries from optical
surveys were 2 days for all observatories, then about 270 NEAs would
have been detectable at Arecibo, or 17 more than with the current
reporting delays. Of the 17 additional targets, 10 are in the highest
SNR category.

As we mentioned in section~\ref{sec:methods}, Arecibo sometimes
targets NEAs with SNRs $< 30$/track during survey nights. If we adopt
a threshold of SNR/track $\geq 20$, then 290 NEAs become detectable at
Arecibo, i.e., a 12\% increase over the number of detectable asteroids
with SNRs $\leq 30$/track.

\subsection{DSS-14 monostatic}

We found 246 objects that were potentially detectable at DSS-14 with
SNRs/track $\geq 30$ in the year 2015
(Figure~\ref{fig:mainhist}). There were 131 discoveries that were
announced on the MPC website before the radar view periods
ended.
DSS-14 observed 32 NEAs or about 24\% of these objects. An additional
seven NEAs with weaker SNRs were also observed.  The targets are
roughly evenly spread among the three categories defined in the
previous section: 58 targets had SNRs/track between 30 and 100 (10
were detected), 34 had SNRs/track between 100 and 300 (2 were
detected), and 39 had SNRs/track greater than 300 that were suitable
for high-resolution imaging (Figure~\ref{fig:snrbins}) (20 were
detected). As we found with Arecibo targets, the time between the
discovery announcement and the end of the radar window was less than
15 days for most.

Recently an engineering study (L. Teitelbaum, pers. comm.) was
conducted to examine the efficacy of designing and building new
klystron amplifiers that would double the transmitter power at DSS-14
from 450 kW to 900 kW.
\textbf{If this upgrade occured,} then the number of detectable NEAs
would have been 165, which is a 26\% increase from the current
capability. Doubling the transmitter power will also increase the
effective range of DSS-14 by 19\%, double the SNRs, and allow us to
image targets with $\sim 2$ times finer resolution than is possible
with current SNRs.

\subsection{Bistatic DSS-14 to Green Bank Telescope}

The number of NEAs detectable by using DSS-14 to transmit and GBT to
receive is greater than the NEAs detectable by monostatic DSS-14
observations but less than the number of NEAs detectable with
monostatic observations at Arecibo. This is correlated to the
collecting area of the receiving telescopes. There were 327 objects
that were potentially observable using DSS-14 and GBT in 2015
(Figure~\ref{fig:mainhist}) and of these 195 were discoveries that
were announced before the end of their radar view periods. This is an
increase of 64 radar-detectable asteroids, or about 50\%, compared to
monostatic observations at DSS-14. The number of NEAs in the highest
SNR category was more than double relative to
DSS-14 monostatic observations (Figure~\ref{fig:snrbins}), indicating
that many of the low and medium SNR objects were promoted to the high
SNR category. In 2015, we observed 8 NEAs using DSS-14 and GBT. This
bistatic configuration was only rarely used for observing NEAs before
2015.

All the objects detectable by DSS-14 monostatically were also
detectable using the DSS-14-GBT bistatic configuration mainly due to
the similar declination ranges covered by the telescopes. The
estimated SNRs were higher in the bistatic configuration because the
Green Bank Telescope has a larger effective aperture (100 m vs. 70 m)
and the total integration time is about two times greater. On average,
the Goldstone (DSS-14) - GBT bistatic configuration provides
SNRs/track about 3 times greater than the Goldstone monostatic
configuration. Note that this factor compares SNRs/track and is
different from the relative sensitivity listed in
Table~\ref{tab:relsens}, which compares SNRs/RTT.

Lowering $T_{sys}$ at GBT would further increase the relative
sensitivity and enable more NEA detections. For example, if $T_{sys}$
were 18 K at GBT, then DSS-14 and GBT \textbf{would have been able to detect
230 NEAs,
an 18\% increase relative to the current capability with $T_{sys}=25$
K.}
A preliminary study shows that lowering $T_{sys}$ to 18 K at GBT is
feasible (J. Ford, personal communication).  \textbf{No such upgrade
  is currently planned.}

There are some limitations to using the DSS-14-GBT bistatic
configuration: the GBT is not always available, the latitude and
longitude differences means that overlapping time is short for targets
at southern declinations,
and weather in the winter can preclude observations.

\subsection{Bistatic Arecibo to Green Bank Telescope}

We found 291 NEAs that were potentially observable with SNRs/track
$\geq 30$ in 2015 using the Arecibo-GBT bistatic configuration
(Figure~\ref{fig:mainhist}). The discoveries of 157 of these
were reported before the end of their radar view periods. This is
fewer than those detectable using DSS-14 and GBT because that
configuration provides longer observing windows and results in higher
SNRs/track compared to the Arecibo-GBT bistatic configuration for some
targets.  The number of detectable asteroids is also lower than that
for monostatic observations at Arecibo because the GBT has a smaller
collecting area than Arecibo.  However, the Arecibo to GBT bistatic
configuration is very useful because it provides longer integration
times and allows for finer Doppler resolution of the data. This is
ideal for very close targets with RTT $< 10$ seconds and/or very slow
rotators with intrinsically narrow bandwidths. Three asteroids (2015
HM10, 2015 SZ, and 2003 SD220) were observed in this configuration in
2015.

\subsection{Bistatic DSS-13 to Green Bank Telescope}

We found 57 near-Earth asteroids that were potentially detectable with
SNRs/track $\geq 30$ in 2015 using this configuration, and of these,
the discoveries of 33 were reported before the end of their radar view
periods. Four asteroids were probably strong enough for delay-Doppler
imaging with the finest possible range resolution of 1.875 m, namely
(357439) 2004 BL86, 2015 HD1, 2015 HM10, and 2015 JF1. We observed 2004
BL86 and 2015 HM10 using this bistatic configuration but we were not
able to achieve the maximum range resolution of 1.875 m due to
technical difficulties.

NEA (85989) 1999 JD6 was the largest detectable asteroid with a
diameter of about 1.8 km and SNR/track of $\sim 400$. It was detected
using DSS-14, Arecibo, and the GBT in 2015 July. The smallest
detectable asteroid was 2015 FM118 (SNR/track $\approx 700$) with an
inferred diameter of 5 m based on an absolute magnitude of 28.7. This
asteroid approached Earth within 0.002 au in March. At this distance
the round-trip light time to the asteroid would have been shorter than
the transmit-receive switching time at Arecibo and DSS-14, so bistatic
observations would have been the only way to detect 2015 FM118 with
radar at the closest approach. The median of the absolute magnitudes
of the detectable asteroids was 25.5, which corresponds to a diameter
of $\sim 24$ m.

\subsection{Bistatic DSS-43 to Parkes}

About 70 NEAs were potentially detectable with SNRs/track $\geq 15$ in
2015 using the bistatic DSS-43 - Parkes configuration and the
discoveries of 34 were reported before the end of their radar view
periods. Table~\ref{tab:dss43parkes} lists all the NEAs that were
observable using this configuration in 2015. One of these asteroids,
2015 BP509, was not observable by any other radar-capable telescopes
considered in this paper because its close approach was at a southern
latitude of -36 degrees that was beyond the reach of telescopes
located in the northern hemisphere.  \\
\begin{deluxetable*}{lllr}
  \tablecaption{NEAs detectable in 2015 using DSS-43 and Parkes
  \label{tab:dss43parkes}}
  \tablehead{\colhead{Object} & \colhead{Absolute magnitude} & \colhead{Distance (au)} & \colhead{SNR/track}}
  \startdata
\textbf{(33342)  1998 WT24}  &	 17.9	&	0.0280 &	16        \\	
(357439)  2004 BL86  &	 19.3	&	0.0080 &	1700      \\	
\textbf{(413577)  2005 UL5}   &	 20.3	&	0.0153 &	87       \\	
(436724)  2011 UW158$^*$ &	 19.9	&	0.0199 &	62        \\	
          2014 YD15  &	 26.8	&	0.0049 &	20        \\	
          2014 YE42  &	 23.4	&	0.0110 &	16        \\	
          2015 BC    &	 24.0	&	0.0050 &	300       \\	
          2015 CL13  &	 25.7	&	0.0054 &	44        \\	
          2015 DS53  &	 24.2	&	0.0081 &	37        \\	
          2015 DY198 &	 26.6	&	0.0056 &	30        \\	
          2015 EF    &	 26.8	&	0.0065 &	16        \\	
          2015 FW117 &	 22.7	&	0.0092 &	72          \\	
          2015 FM118 &	 28.7	&	0.0031 &	70         \\	
          2015 GU    &	 28.4	&	0.0037 &	39         \\	
          2015 GL13  &	 28.8	&	0.0041 &	23         \\	
          2015 HD1   &	 27.4	&	0.0004 &	380000    \\	
          2015 HM10  &	 23.6	&	0.0039 &	590       \\	
          2015 HQ11  &	 27.1	&	0.0053 &	20        \\	
          2015 HO116 &	 25.5	&	0.0045 &	150       \\	
          2015 HQ171$^*$ &	 26.9	&	0.0052 &	26        \\	
          2015 HA177$^*$ &	 27.7	&	0.0048 &	24         \\	
          2015 KW120 &	 26.0	&	0.0035 &	300       \\	
          2015 KA122 &	 23.2	&	0.0085 &	63        \\	
          2015 LF    &	 26.6	&	0.0014 &	6000      \\	
          2015 OQ21$^*$  &	 27.9	&	0.0040 &	32         \\	
          2015 SZ2$^*$   &	 25.4	&	0.0034 &	700       \\	
          2015 TC25$^*$  &	 29.5	&	0.0013 &	1600       \\	
          2015 TB145 &	 20.0	&	0.0078 &	4400         \\	
          2015 VO142$^*$ &	 29.0	&	0.0026 &	130        \\	
          2015 XP    &	 25.8	&	0.0038 &	210       \\	
          2015 XX128$^*$ &	 26.0	&	0.0062 &	24        \\	
          2015 XR169$^*$ &	 28.7	&	0.0035 &	35         \\	
          2015 XY261 &	 27.2	&	0.0019 &	2100      \\	
          2015 YQ1$^*$   &	 28.1	&	0.0038 &	39         \\	
          \enddata
          
          \tablecomments{Distance indicates the minimum distance of the
            asteroid from Earth within the radar observing
            window. Objects in \textbf{bold} were observed by DSS-43
            and Parkes. Asterisk ($^*$) indicates that the asteroid is
            on the NHATS list.}
          
  \end{deluxetable*}

NEA (33342) 1998 WT24 was the largest detectable asteroid in this
configuration with a diameter of $\sim 400$~m (Busch et al. 2008). The
maximum SNR/track for this object was about 10-20 in 2015 December,
when it approached Earth at a distance of 0.03 au. It was the second
asteroid detected by DSS-43 and Parkes after NEA 2005~UL5. The
smallest detectable asteroid was 2015~TC25, with a diameter of 3 m
inferred from its absolute magnitude of 29.5. It approached within
0.001 au and its maximum SNR/track was about 1500. This asteroid was
detected by Arecibo in 2015 October, when it was 0.01 au from Earth or
about 10 times further away than it was at closest approach. The
median of the absolute magnitudes of all detectable asteroids by
DSS-43 and Parkes was 26.6, corresponding to a diameter of 14 m.

Currently the transmitter power at DSS-43 is restricted to less than
100 kW, however it could potentially be raised to 400 kW in the
future. \textbf{With the higher power 62 NEAs (an increase of 82\%)
  would have been detectable using DSS-43 and Parkes.} Transmission at
higher power would require radiation clearance from government
authorities such as local air traffic control, which will increase the
lead time required for scheduling radar observations at DSS-43.

\subsection{Absolute magnitude/size distribution of radar-detectable asteroids}

\begin{figure*}
  \plotone{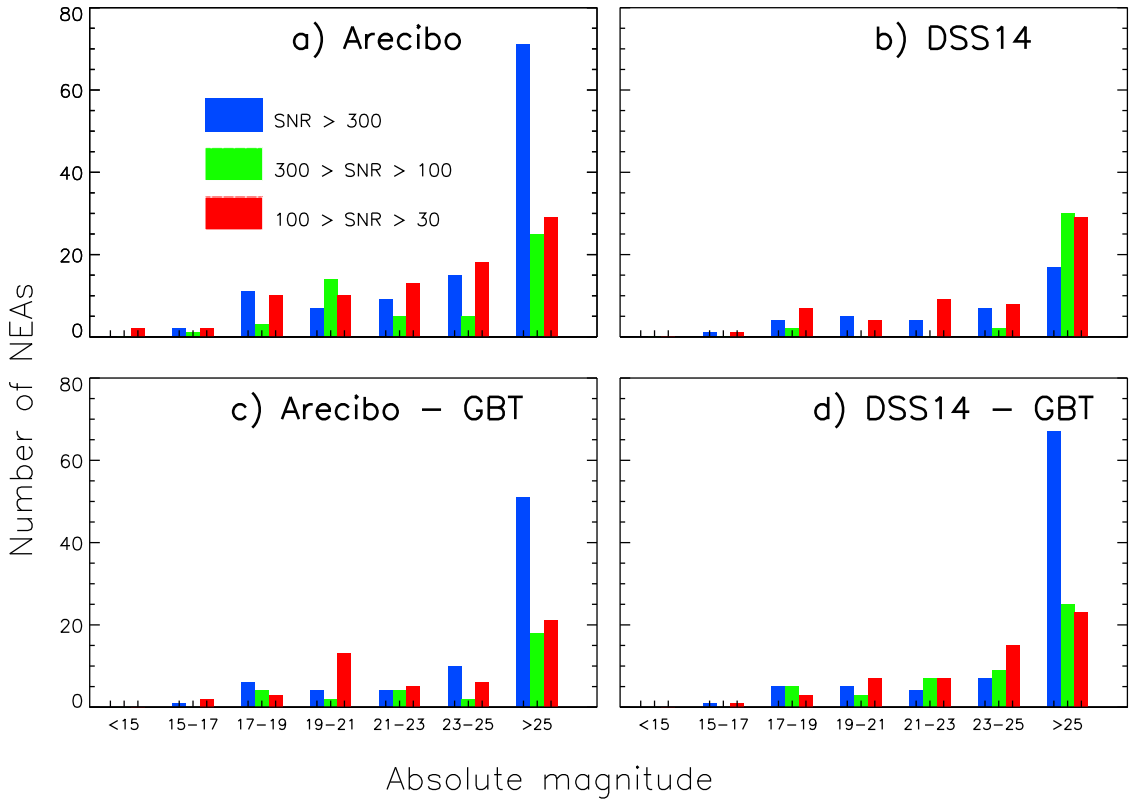}
  
  \caption{Histogram of the number of radar-detectable NEAs as a
    function of absolute magnitude for Arecibo (top left), Goldstone
    (top right), Arecibo-GBT (bottom left) and Goldstone-GBT (bottom
    right). The targets are divided into high SNR (blue), medium SNR
    (green), and low SNR (red) categories. }

  \label{fig:abs}
\end{figure*}

Figure~\ref{fig:abs} shows the distribution of absolute magnitudes of
the radar-detectable asteroids for each transmitter-receiver
configuration. It includes only those objects whose discoveries were
announced before their radar view periods ended. For a typical S-class
asteroid optical albedo of 0.18, magnitudes of 15, 20, and 25
correspond to diameters of about 3000 m, 300 m, and 30 m respectively
but the size of a given object may vary by a factor of two due to
albedo assumptions.

The smallest radar-detectable asteroid was 2015 VU64, with an absolute
magnitude of 30.6 and a diameter of $\sim 2$~m. It approached Earth to
within 0.0007 au (16.4 Earth radii) and was observable only with a
bistatic configuration because the round-trip light time to the
asteroid during the radar tracks was less than the transmit-receive
switching time at all telescopes. The maximum SNR/track was over
$10^6$. The largest detectable asteroid was (152679) 1998 KU2, which
is $\sim 4.6$~km in diameter and approached within 0.25 au. It was
detectable only at Arecibo.

The median of the absolute magnitudes of all radar detectable
asteroids is 25.4, which corresponds to a diameter of $\sim 25$~m. For
Arecibo, most of the asteroids fainter than 25 mag are in the high SNR
category, but for Goldstone medium and low SNR targets constitute the
majority of the small asteroids. Figure~\ref{fig:abs} shows that using
the GBT as a receiver with DSS-14 promotes all small targets from the
medium to the high SNR category.

Objects $\sim25$~m in diameter are worth tracking with radar due to
the impact risk such as the Chelyabinsk airburst in
2013~\citep{popova13}, because some will be mission targets, and due
to scientific interest in their physical properties.

Out of 1562 NEAs discovered in 2015, $\sim 66\%$ had absolute
magnitudes $> 22$ (diameter $< 140$ m). This explains the high
fraction of small radar-detectable asteroids. The means $\pm$ standard
deviations of the absolute magnitudes of the asteroids observed by
Arecibo and DSS-14 are $21.1\pm3.2$ and $20.9\pm3.3$ respectively.
This implies that radar detections are biased towards observing
optically brighter asteroids.
\textbf{This bias is partly due to the brighter/larger targets being
  prioritized as a result of limited resources, such as telescope time
  and observing staff, but also due to time available to prepare for
  the observations.}

\section{Discussion}

Out of the 13513 known NEAs at the end of 2015, 430 and 246 had SNRs
high enough to be detected by Arecibo and Goldstone, respectively,
during the calendar year. About 40-50\% of these were not observable
because the discoveries were reported after the end of their radar
observing windows leaving 253 and 131 targets that were actually
detectable at Arecibo and Goldstone respectively. Of these, Arecibo
and Goldstone observed 30\% (77 out of 253) and 24\% (32 out of
131). Combined, the two observatories were capable of observing 276
NEAs in 2015. This number is different from the corresponding value of
410 reported in Shapiro et al. (2010) because we adopted a SNR
threshold of 30/track versus 5/track in the earlier study. The
analysis in the current study is much more rigorous because we
computed detailed SNRs for each object based on the actual close
approach circumstances.

Shortening the delay in reporting NEA discoveries from optical surveys
would increase the potential of radar observatories. If all the
optical surveys reported discoveries within 2 days, then about 17
additional objects, including 10 high SNR objects, become detectable
at Arecibo.

The bistatic DSS-14 to GBT configuration increases the number of
radar-detectable targets by about 50\% compared to the monostatic
DSS-14 configuration.
\textbf{Most of the low and medium SNR targets in the monostatic case
  are promoted to the high SNR category with reception at the GBT.
  Some of these high SNR targets that were not observed could have
  been imaged with resolutions as fine as 3.75~m.}

Doubling the transmitter power at DSS-14 increases the number of
detectable NEAs by about 26\%. In an earlier different study,
\citet{giorgini08} reported that a 5\% increase in detectable NEAs
would be achieved by doubling the transmitter power at DSS-14. The two
studies use different approaches in the sense that we looked at known
NEAs whereas \citet{giorgini08} looked at a simulated NEA population.
The two studies adopted different SNR thresholds: 30/track in this
study vs. 10/track in \citet{giorgini08}.  The methodology was
different in the two studies because they had different goals but
their results are consistent. 

This study clearly shows that Arecibo and Goldstone are observing less
than one-half of potentially detectable NEAs. The number of NEAs
observed by radar could be increased by at least several tens of
percent by obtaining more telescope time at Arecibo and Goldstone
without changing protocols to respond more rapidly. Most of the
radar-detectable asteroids leave the detectability window within 5
days of their discovery announcement and have absolute magnitudes $>
25$. For Arecibo and Goldstone to observe these targets, a more rapid
response time is necessary at both telescopes and a higher level of
dynamic scheduling and staffing is needed that would allow switching
quickly from previously scheduled observations to active radar
observations.  More rapid discovery announcements would allow more
lead time to prepare for radar observations thereby enabling radar
detection of more asteroids.

\acknowledgements

This research was conducted at the Jet Propulsion Laboratory,
California Institute of Technology, under contract with the National
Aeronautics and Space Administration (NASA). The material presented
represents work supported by NASA under the Science Mission
Directorate Research and Analysis Programs. Part of the work was done
at the Arecibo Observatory, which is operated by SRI International
under a cooperative agreement with the National Science Foundation
(AST-1100968) and in alliance with Ana G. Mendez-Universidad
Metropolitana and the Universities Space Research Association. The
Arecibo Planetary Radar Program is supported by the National
Aeronautics and Space Administration under Grant Nos. NNX12AF24G and
NNX13AQ46G issued through the Near Earth Object Observations program.
\textbf{We thank the staff at the Deep Space Network and Arecibo. We
  thank Jon Giorgini for many useful discussions and for comments on
  earlier versions of this manuscript.}

\pagebreak


\end{document}